\documentclass[sigconf]{acmart}

\usepackage{balance}

\def\BibTeX{{\rm B\kern-.05em{\sc i\kern-.025em b}\kern-.08emT\kern-.1667em\lower.7ex\hbox{E}\kern-.125emX}}

\begin{document}
\nocite{*}

\copyrightyear{2019} 
\acmYear{2019} 
\setcopyright{acmcopyright}
\acmConference[KDD '19]{The 25th ACM SIGKDD Conference on Knowledge Discovery and Data Mining}{August 4--8, 2019}{Anchorage, AK, USA}
\acmBooktitle{The 25th ACM SIGKDD Conference on Knowledge Discovery and Data Mining (KDD '19), August 4--8, 2019, Anchorage, AK, USA}
\acmPrice{15.00}
\acmDOI{10.1145/3292500.3330707}
\acmISBN{978-1-4503-6201-6/19/08}

\settopmatter{printacmref=true}
\fancyhead{}
%
\title{Real-time Attention Based Look-alike Model for Recommender System}

%
\author{Yudan Liu}
\affiliation{%
  \institution{WeiXin Group, Tencent Inc.}
  \city{Beijing}
  \state{China}
}
\email{danydliu@tencent.com}

\author{Kaikai Ge}
\affiliation{%
  \institution{WeiXin Group, Tencent Inc.}
  \city{Beijing}
  \state{China}
}
\email{kavinge@tencent.com}

\author{Xu Zhang}
\affiliation{%
  \institution{WeiXin Group, Tencent Inc.}
  \city{Beijing}
  \state{China}
}
\email{xuonezhang@tencent.com}

\author{Leyu Lin}
\affiliation{%
  \institution{WeiXin Group, Tencent Inc.}
  \city{Beijing}
  \state{China}
}
\email{goshawklin@tencent.com}



%

%
\begin{abstract}
Recently, deep learning models play more and more important roles in contents recommender systems. However, although the performance of recommendations is greatly improved, the "Matthew effect" becomes increasingly evident. While the head contents get more and more popular, many competitive long-tail contents are difficult to achieve timely exposure because of lacking behavior features. This issue has badly impacted the quality and diversity of recommendations. To solve this problem, look-alike algorithm is a good choice to extend audience for high quality long-tail contents. But the traditional look-alike models which widely used in online advertising are not suitable for recommender systems because of the strict requirement of both real-time and effectiveness. This paper introduces a real-time attention based look-alike model (RALM) for recommender systems, which tackles the challenge of conflict between real-time and effectiveness. RALM realizes real-time look-alike audience extension benefiting from seeds-to-user similarity prediction and improves the effectiveness through optimizing user representation learning and look-alike learning modeling. For user representation learning, we propose a novel neural network structure named attention merge layer to replace the concatenation layer, which significantly improves the expressive ability of multi-fields feature learning. On the other hand, considering the various members of seeds, we design global attention unit and local attention unit to learn robust and adaptive seeds representation with respect to a certain target user. At last, we introduce seeds clustering mechanism which not only reduces the time complexity of attention units prediction but also minimizes the loss of seeds information at the same time. According to our experiments, RALM shows superior effectiveness and performance than popular look-alike models. RALM has been successfully deployed in "Top Stories" Recommender System of WeChat, leading to great improvement on diversity and quality of recommendations. As far as we know, this is the first real-time look-alike model applied in recommender systems.
\end{abstract}

%
%


%

%
\maketitle

\section{INTRODUCTION}
As for contents recommender systems, the traditional recommendation algorithms such as collaborative filtering\cite{sarwar2001item} and content based algorithms\cite{pazzani2007content} have been applied widely. Recently, deep learning models such as deep neural networks (DNNs) and recurrent neural networks (RNNs) are more and more popular on recommendation task\cite{covington2016deep}\cite{cheng2016wide}\cite{chen2017attentive}. These deep learning based methods effectively capture the user preferences, item features and non-liner relationship between user and item, which show better performance compared with traditional algorithms on recommendation in most situations.

\begin{figure}[h]
    \centering
    \includegraphics[scale=0.68]{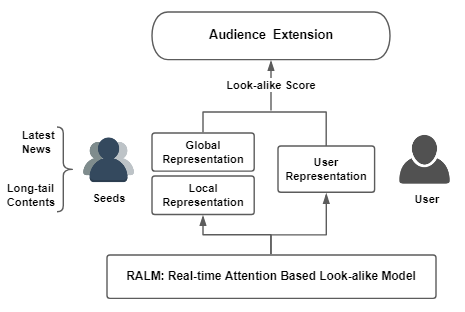}
    \caption{Audience Extension in Recommender System}
    \label{intro}
\end{figure}

However, these end-to-end models like deep learning networks aim at improving the performance of recommendation, and tend to predict higher CTR (click-through rate) for the head contents than the long-tail ones. It means that popular articles and those target user have clicked are always preferred. At the same time, there are many competitive long-tail contents including contents from manual pushing, novelties and latest news. These long-tail contents are usually short of behavior features, which are essential for recommendation models. As a result, they are difficult to achieve wide and timely exposure. We call it the "Matthew effect" in recommender systems, leading to low quality and poor diversity of recommended contents. Apart from the performance, improving the quality and diversity of recommendation results have become common problems faced with many recommender systems.

To solve this problem, we need an efficient user targeting strategy helping to extend the audience of those good long-tail contents. Look-alike modeling\cite{mangalampalli2011feature} is popular on audience extension technology which is widely used in online advertising\cite{liu2016audience}. As there is a candidate for audience extension and a certain amount of users who have clicked it, named seeds, the look-alike models essentially aim to find group of users who are similar to the seeds. Because of less features for long-tail contents in recommendation campaign, the look-alike model is a good choice which only depends on the seed users as inputs instead of specific features of contents. However, unlike advertising, the efficacy of audience extension in recommender system is measured by multiple aspects:
\begin{itemize}
\item \textbf{Real-time} - The articles which need audience extension are produced continually by news pipeline and manual operations. They have to be distributed in real-time owing to the strict time sensitivity. This is the most important feature of audience extension in recommender system.
\item \textbf{Effective} - The audience extension is a complementary service independent of existing recommendation framework based on CTR prediction. We have to improve the effectiveness and try the best to preserve the performance on CTR prediction. At the same time, it requires high accuracy and diversity of user interest representation and seeds feature representation.   
\item \textbf{Performance} - There are tens of thousands candidates available for audience extension at the same time and up to millions of seeds for one candidate. The audience extension system should score all candidates with high performance online. So that the look-alike model can not be too complicated due to time complexity.
\end{itemize}

The traditional look-alike models such as regression based models\cite{tufekci2008can} have to be trained against each candidate offline, then do prediction online. Therefore they can not be real-time. The similarity based look-alike algorithms make real-time work, but the effectiveness is much inferior to the regression models. To solve this problem, Yahoo came up with hybrid model in 2016\cite{ma2016sub}. This model actually balances the performance and effectiveness through a user-to-user similarity graph, but it still need several hours of time to prepare the feature weights for each candidate before coming into force. Generally speaking, real-time look-alike models are usually based on seeds-to-user similarity calculation, which may lead to the damage of accuracy because of poor representation of user and seeds, and the difficulties are listed in detail as following:
\begin{itemize}
\item  \textbf{User Representation.} To improve the diversity of user interest, as many as fields of user features should be used for user representation learning.  The deep learning based models are capable to model multiple fields of features. However, according to our experiments, the deep models like DNN show worse performance if feeding weak-relevant fields of features as input. We believe the strongly-relevant features like interested tags lead to overfitting, while the weakly-relevant features such as shopping interests take limited even negative effect. 
\item \textbf{Seeds Representation.} The seeds are accumulating continually in recommender system and will consist of various and a large amount of users, even some noisy users. How to represent the seeds using the representation of each user is another difficult task. For one thing, to improve the robustness, every seed should take different contribution to the seeds group. For another, due to large amount of users in the seeds, the target user may only be similar to a part of them. So we should model the local information to improve the adaptivity.
\end{itemize}

In this paper, a approach named Real-time Attention Based Look-alike Model(\textbf{RALM}) is designed to not even realize real-time look-alike model, but also guarantee the effectiveness. RALM is a similarity based look-alike model, which consists of user representation learning and look-alike learning.  \textbf{(1).} For user representation learning, instead of  applying traditional concatenate layer, we design a novel network structure named attention-merge layer which shows significant superior performance on multi-fields features. \textbf{(2).} To optimize the seeds representation, look-alike learning models global representation and local representation of seeds based on global and local attention units. \textbf{(3).} RALM applies seeds clustering which is processed asynchronously online to reduce the scale of seeds and the time complexity of online prediction on attention units. The main contributions of our work are summarized as follows: 
\begin{itemize}
\item \textbf{Improve the effectiveness of user representation learning.} We design a specific deep interest network for multi-fields user interests representation learning, which introduces a novel network structure named attention merge layer. The attention merge layer solves overfitting and noise problems brought by strongly-relevant features and weakly-relevant features. According to our experiment, the attention merge layer is much more effective to capture various fields of interests compared with concatenation layer which is widely used now.
\item \textbf{Improve the robustness and adaptivity of seeds representation learning.} \textbf{(1).} We apply global attention unit to learn the global representation of seeds. Global attention unit weights the representative individuals and punishes the noisy users, which is much more robust than equally processing. \textbf{(2).} Meanwhile, we introduce local attention unit to extract the local representation of seeds, which weights the relevant individuals to target user. The local attention unit adaptively learns seeds representation with respect to a certain target user, so that the representation of seeds varies over different audiences, which improves the seeds expressive ability significantly.
\item \textbf{Realize a real-time and high-performance look-alike model.} To update the latest seeds information, the global and local representation of seeds should be processed online in real-time. Considering the large amount of computation on attention units, we come up with k-means clustering to partition seeds into k clusters. This novel process contributes that the complexity of look-alike computation is greatly reduced while minimizing the loss of seeds information. Meanwhile, because the clustering result changes while the vector of seeds fine tuned in training, we introduce an iterative training progress of seeds clustering and deep model training. Furthermore, based on the seeds-to-user look-alike method, only the vectors of seeds and target user are fed as input of the predicting model, so the candidate articles go into effect as soon as seeds uploaded. As far as we know, this is the first real-time look-alike model applied in recommender system.
\end{itemize}

\section{BACKGROUND}
WeChat is the most popular messaging app in China, which has more than 1.4 billion users. Except messaging, WeChat also provides contents services. Users are able to reach personalized articles, news and videos from "Top Stories", a product of contents feed powered by personalized recommender system. There are millions of new contents produced everyday and billions of daily page view. As we introduced in last chapter, due to the "Matthew effect", an effective audience extension is what we need to promote the recommendations of the competitive long-tail contents.

\section{RELATED WORK}
To implement audience extension, recent works mainly focus on user representation and look-alike algorithm.

User representation purposes on representing a user in an effective form based on their features. The easiest way is to denote a user by a feature vector $f=(f_{1},···,f_{n})$, in which $f_{n}$ is a categorical feature\cite{shen2015effective}. In many cases, the dimension of a feature vector can be up to millions and sparse, resulting as an inefficient representation. LSH is another way of user representation. The dimension of feature vectors is reduced, and each user is assigned to a cluster based on the hash signature, or alternatively, assigned by clustering algorithm like K-means\cite{hartigan1979algorithm}. However, these methods make calculations directly on raw feature values, and no iterative learning task is used. This representation is coarse and lack of information.

Deep learning models are getting more and more popular in recommender systems because of their ability to learn high-level interactions and hidden information among variety of features. Different from CTR prediction, deep learning models in user representation focus on learning a dense "embedding" vector for users. Youtube DNN model\cite{covington2016deep} shares a widely used deep network structure to learn the embedding of user. However, with regard to multi-fields user features, including both strong and weak features, we draw a conclusion that concatenation layer leads to bad performance because of overfitting on strong features and underfitting on weak features.


As to look-alike algorithm, previous works are mainly concluded into two categories: similarity based methods and regression based methods.

Similarity based methods determine similarity between seeds and users based on distance measurement. With user embedding vectors learned from user representation tasks, the similarity between two users can be measured based on their embeddings. Jaccard similarity, Cosine similarity and dot product are available measures. Then the similarity between set of users (seeds) and a target user can be calculated as average similarity of the target user to all of the seeds. In this way, users are simply targeted to majority of users in seeds, and information carried by less but possibly more similar users is lost.

Regression based methods treat look-alike as a regression problem. A simple but impressive method is applying LR(Logistic Regression) model\cite{qu2014systems} to train a regression model against each target item. Seed users will be treated as positive example, and sample a certain amount of non-seed users to be negative example. As a result, users who look like the seeds will be predicted with higher score. In addition, FM\cite{rendle2010factorization} and MLP (Multiple Layer Perceptron)\cite{kim2006performance} have been used in audience extension. All of these regression based models essentially aim to maximize the possibility of observed behavior of seeds based on user features. But the problem is the training process must take much time offline against every target item. On the one hand, regression based models need accumulating seeds as positive examples to update the current model. On the other hand, the model has to be trained again if a new item is added. When new items are frequently generated in the system, they are hard to come into force in time. So regression based models can not be applied on real-time audience extension. 

Besides, Yahoo posted a method combining above two methods\cite{ma2016sub}. There are two steps in this method. First, all users are clustered, and a candidate user set is generated according to the article. Second, use a regression based model like LR or even a simple feature selection to filter out irrelevant users. This method is designed to adapt massive dataset and large-scale audience extension in Yahoo. Meanwhile, it abandons complexity and precision of look-alike algorithm as a compromise of online performance. 

Attention mechanism is firstly applied in Machine Translation\cite{bahdanau2014neural}. It supposes that when faced with a sentence, man pays different attention to the words according to different context. General attention\cite{vaswani2017attention} outputs a weight matrix, by which word embeddings can be weighted summed to one according to a context embedding. Self attention\cite{lin2017structured} takes only current words as consideration, and learn weights to sum with themselves. Recently, attention has been applied in recommender systems successfully, such as DIN\cite{zhou2018deep} which proved that attention mechanism is able to extract diverse user interests on variety of features.

\section{SYSTEM}

\begin{figure*}[h]
    \centering
    \includegraphics[scale=0.46]{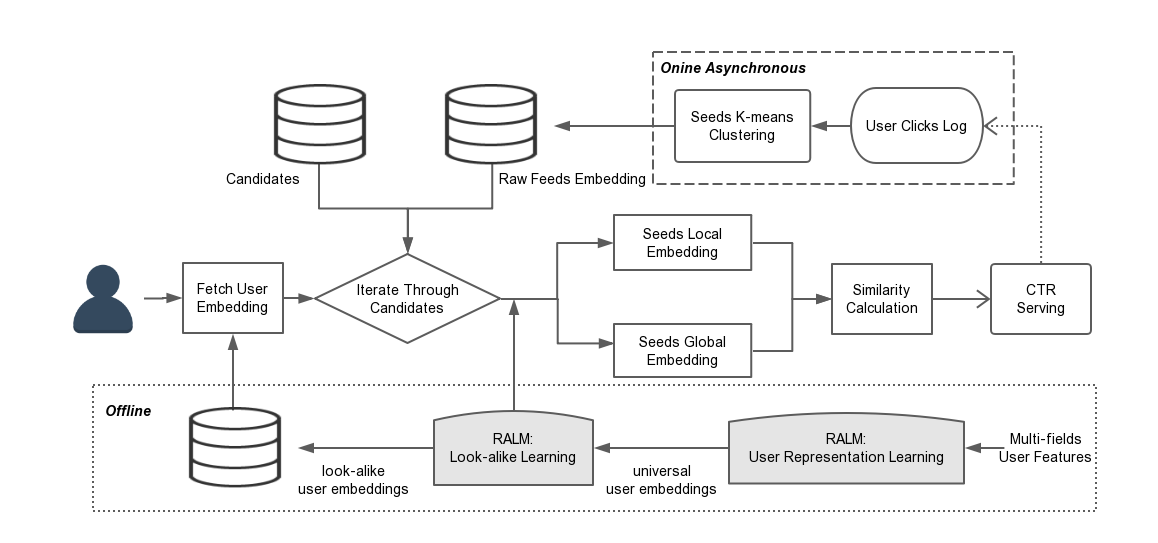}
    \caption{RALM Based Audience Extension System Framework and Pipeline}
  \label{framework}
\end{figure*}
In this section we will introduce the framework of RALM based audience extension system from high-level perspective and more details will be presented in next section from modeling perspective. Fig.2 shows the overview of framework and pipeline of RALM system.

\subsection{Overview}
There are several types of candidates for audience extension in "Top Stories" recommender system of WeChat, such as latest news, manually-tagged high quality articles and long tail interest contents. All of them are produced through real-time pipeline, updating the candidates database online. There are thousands of candidates available for audience extension at the same time. For each available candidate, the system collects the seeds who click it asynchronously to update the seeds-cluster embedding database.

The user vectors are produced offline by user representation learning, while the local and global vectors of seeds are predicted online based on the vectors of seed clusters and the offline look-alike model. When an audience creates a request, serving process will firstly fetch the vector of the current audience, and iterate through the candidates to score the look-alike similarity between audience and seeds of candidates.
    
The key components can be concluded into three parts: offline training, online asynchronous processing, online serving.
    
\subsection{Offline Training}
The online serving of audience extension depends on the user embeddings and the model for seeds embedding. We propose to train look-alike model offline in two phases, named user representation learning and look-alike learning. 
	
\begin{itemize}
\item Phase1: User Representation learning. The user representation model is developed based on deep learning network, which takes all fields of user features as input and user behaviors in WeChat as training samples, such as reading articles, playing videos, shopping goods, playing musics, subscriptions and so on. The output of user representation model is the universal user embeddings which represent multi-fields of features.
\item Phase2: Look-alike learning. Look-alike learning is based on attention model and clustering algorithm, which takes universal user embeddings generated in last phase as input and audience extension campaign samples as training samples, producing user embeddings for look-alike which is used for look-alike similarity prediction. On the other hand, the attention models of global and local seeds embeddings are generated in this process, which are used for predicting adaptive seeds representation.
\end{itemize}

\subsection{Online Asynchronous Processing}
Online asynchronous processing mainly aims at updating seeds embedding database in real-time. The members of seeds are accumulating while audience extension serving, and as we introduced in introduction section, we apply K-means clustering to partition seeds into k clusters. So the asynchronous workflow consists of two components:

\begin{itemize}
\item User feedback monitor: The audience extension system updates the seeds of candidates through monitoring the click behaviors of all WeChat users in real-time. Considering explosive growth of seeds may impact the performance of clustering, we keep the latest 3 million click-users as the seeds of a candidate.
\item Seeds clustering: Although the seeds are updated in real-time, clusters are not necessary updated every time when new seed inserted in. The system runs K-means algorithm every 5 minutes to cluster the new seeds. The embeddings of cluster centroids are saved in database as the raw representation of seeds, which will be used for online predicting of seeds embeddings. The raw representation of seeds $R_{seeds}$ is defined as:
\begin{equation}
	\quad \quad \quad R_{seeds}=\left \{  \right.E_{centroid_{1}},E_{centroid_{2}},\cdot \cdot \cdot ,E_{centroid_{k}}\left.  \right \},
\end{equation}
where $E_{centroid_{k}}$ is the embedding of centroid of the k-th cluster.
\end{itemize}

\subsection{Online Serving}
Firstly, audience extension system fetches the look-alike embedding of current user. Then for each candidate, we fetch the centroid embeddings of the seeds as input of look-alike model. The look-alike model predicts the global embedding of seeds through global attention unit, and predicts the local embedding of seeds through local attention unit. The attention units will be described in detail next section. Finally, online serving module scores the look-alike model by calculating the global similarity and the local similarity. Given user $u$ and seeds $s$, the score of look-alike model is defined as:
\begin{equation}
	score_{u,s}=\alpha\cdot cosine(E_{u},E_{global_{s}})+\beta \cdot cosine(E_{u},E_{local_{s}}),
\end{equation}
where $E_{global_{s}}$ is the global embedding of seeds, $E_{local_{s}}$ is the local embedding of seeds, $\alpha$ and $\beta$ are the weight factors for the global and local similarity. We adapt $\alpha=0.3$ and $\beta=0.7$ for "Top Stories" recommender system of WeChat. The look-alike model score will be taken as weighting factor in ctr prediction work-flow.

Because RALM is based on similarity calculation and only take high-level embeddings as input, the look-alike online serving process is simple and high-performance.

\section{MODEL}
In this section, we will present how we learn user representation, seeds representation and model the look-alike similarity between seeds and target user.

\subsection{Features}
There are many features contributed to user interests including two types: categorical features and continuous features. Categorical features contain univalent (like gender, location) and multivalent (like interested keywords) features. For a value or a set of value representing one kind of categorical features, we call it a feature field. As for continuous features such as age, pre-trained feature vectors are normalized and scaled to the range between 0 and 1. For example, in WeChat, the features include gender, age, location, interested tags, interested categories, apps logged in, media ids, accounts subscribed, shopping interests, search interests, social network relations and so on.

\subsection{User Representation Learning}


User interest can be very complicated and diverse, just like his/her age, nation or what he/she has read that determine what he/she will read the next. Thus we design a deep neural network model involving variety of user features to learn a comprehensive representation of user's interest on contents. User representation learning is elaborated in three parts: sampling, model structure and attention merge layer.


\begin{figure*}[h]
    \centering
    \includegraphics[scale=0.37]{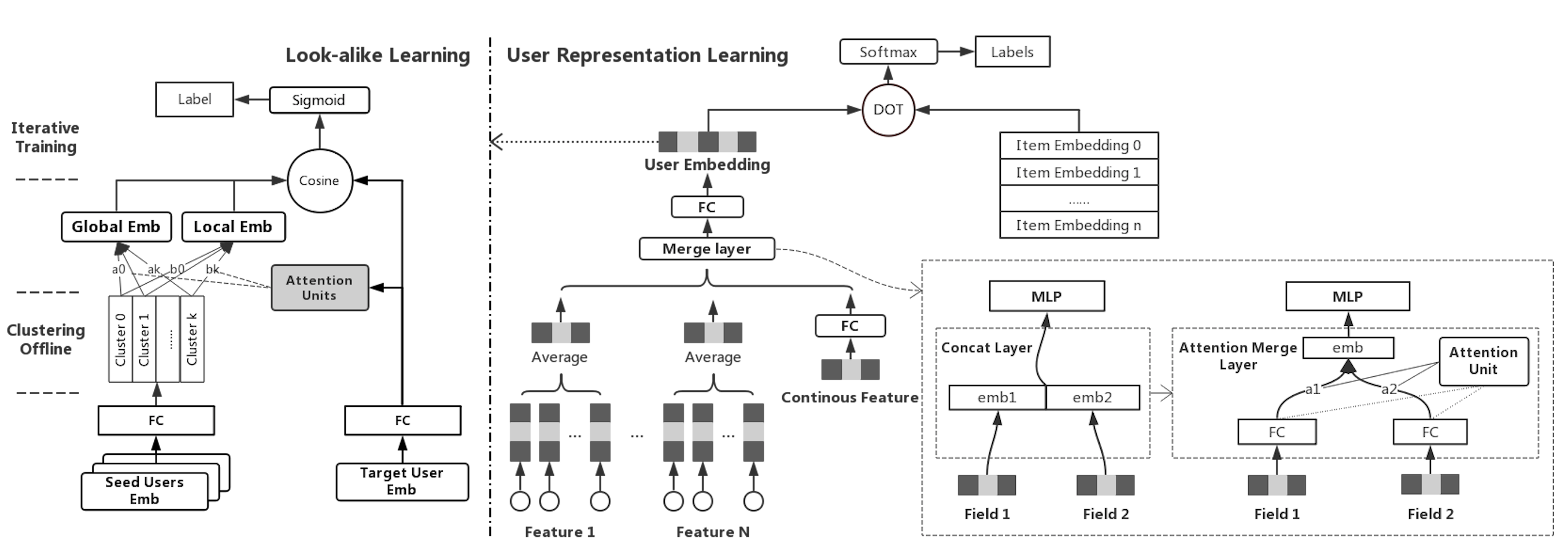}
    \caption{Model Structure of user representation learning. The right part is network of user representation learning, in which concatenation and attention merge layer are listed in comparison. In a mini-batch, features of one user and multiple item embeddings are fed forward the network. The left part is network of look-alike learning, involving an iterative training on clusters.}
    \label{User Representation Learning}
\end{figure*}

\textbf{\textit{Sampling}}. We treat user representation learning as multi-class classification that chooses an interest item from millions of candidates. To improve training efficiency, we use negative sampling instead of traditional softmax when calculating loss. Obviously, if we pick items as negative example randomly, the sampled distribution will deviate from reality. So we developed a loss like NCE loss noted by Google in word2vec\cite{mikolov2013distributed}. To imitate an unbiased distribution, we firstly rank all candidate items by their frequency of appearance, and then calculate the possibility of each item depending on its rank:
\begin{equation}
p(x_i)=\frac{\log(k+2)-\log(k+1)}{\log(D+1)},
\end{equation}
where $x_i$ denotes the i-th item, k denotes the rank of the i-th item, D denotes the max rank of all the items, and $p(x_i)$ represents possibility of picking the i-th item as a negative example. In case that the most active users' behaviors dominate the training loss, we limit max number of positive examples for each user to 50, and sample in a positive/negative proportion of 1/10. Then we apply softmax function to normalize possibilities of a choice $c$ on an item $i$ from those sampled items based on user's feature $U$ and feature of the item $Xi$:
\begin{equation}P(c=i|U, X_i) = \frac{e^{x_{i}u}}{\sum_{j\in{X}}e^{x_{j}u}},
\end{equation}
where $u \in{\mathbb{R}^{N}}$ denotes a high-dimensional embedding of the user, $x_j\in{\mathbb{R}^{N}}$ denotes embeddings of all the items. In this task, user embeddings are produced from a deep neural network, which will be elaborated in the part of model structure. 

In addition, we take both explicit and implicit feedback as example to ensure diversity of recommendation results. Behavior on all types of contents like article, video and website will be considered as example to ensure the representation cover more aspects of user's interest.

\textbf{\textit{Model structure}}. We refer to Youtube DNN as the base model, which includes an embedding layer, a concatenation layer and a MLP layer. In embedding layer, all features of same field are embedded into fixed length vectors, then fed into average pooling layer. After all fields of features are embedded, we concatenate them as a wide dense layer which are fed into MLP layers. At last, the output of the last layer is the embedding of the user. Notice that item embeddings are randomly initialized, and will be updated while training. It is convenient for this model to handle heterogeneous and multi-fields features. With user embedding $u$ and the i-th item embedding $x_i$, we calculate $P(c=i|U, X_i)$ and the cross entropy loss:
\begin{equation}
L = -\sum_{j\in{X}}y_{i}\log{P(c=i|U, X_i)},
\end{equation}
where $y_{i}\in\{0,1\}$ is the label, and we use Adam Optimizer to minimize the loss. When loss converges, the output of last layer will be the representation of user interests.

\textbf{\textit{Attention merge layer}}. In base model, different feature fields are concatenated like $e_{i}=(e_{i_1};e_{i_2};...;e_{i_k})$. However, through observing the training process of network parameters, we find that optimization is always overfitted to some certain fields, which are strongly relevant to user interest on contents, such as interested tags. It leads to a result that the recommended results are determined by the few strongly-relevant fields. However, weakly-relevant fields, which are usually underfitting, such as shopping interests should also make contribution to recommendation. As a result, the model can not learn comprehensively on multi-fields features, and will lack diversity of recommended results. 

To solve this problem, we pose attention merge layer instead of concatenation in representation model. In base model, concatenation forces all users' interest to be learned into the same distribution. In this way, a few strongly-relevant fields affecting majority of users are much weighted, resulting a high-dimensional and sparse weight matrix. Whereas attention unit can learn personalized distribution of weights according to contextual user features, and activate different parts of neurons for various fields. It ensures that strongly-relevant fields and weakly-relevant fields both make effects during training. Therefore we design attention merge layer to learn user-related weights for multiple fields. 

As shown in right part of Fig.3, $n$ fields are embedded with the same length $m$ as vector $h\in{\mathbb{R}^{m}}$, and then we concatenate them in dimension 2, resulting a matrix $H\in{\mathbb{R}^{n\times{m}}}$. Next, we compute weights vector $a$ as follows:
\begin{equation}
	u=\tanh(W_{1}H),
\end{equation}
\begin{equation}
	a_i=\frac{e^{W_{2}u_{i}^{T}}}{\sum_{j}^{n}e^{W_{2}u_{j}^{T}}},
\end{equation}
where $W_{1}\in{\mathbb{R}^{k\times{n}}}$ and $W_{2}\in{\mathbb{R}^k}$ are weight matrix, $k$ denotes size of attention unit, $u\in{\mathbb{R}^{n}}$ represents the activation unit for fields, $a\in{\mathbb{R}^{n}}$ represents weights of fields. Afterwards, we calculate merged vector $M\in{\mathbb{R}^{m}}$:
\begin{equation}
	M=aH.
\end{equation}
Then we take it as the input of the MLP layer and get universal user embedding. Attention merge layer gains a greatly improvement over concatenation, which will be presented in the section of experiments.

\subsection{Look-alike Learning}
Now we have high-dimensional representation namely universal user embedding for all users. What to do next is learning correlation between seed users and target user. This task is as well a supervised learning on user-item behavior, whereas the difference from representation learning is that this task is focused on a specific campaign, on which users just show part of their interest. We will introduce look-alike learning in following parts: model structure, transforming matrix, local attention, global attention, iterative training and loss.


\textbf{\textit{Model structure}}. Look-alike learning model builds up with two towers as the left part in Fig.3. The left tower called "seeds tower", takes embeddings of $n$ seed users as input, denoted as $R_{seeds} \in{\mathbb{R}^{n\times{m}}}$,  where $m$ denotes dimension of user representation embeddings. A fully-connected layer acting as a transforming matrix is presented at first layer, converting input matrix of size $n\times{m}$ to size $n\times{h}$, where $h$ denotes dimension of the transformed embeddings. Afterwards, a self attention unit and a general attention unit are involved in pooling embeddings into one vector of size $h$. Meanwhile at the right tower called "target tower", a vector of size $m$ is simply transformed to size $h$. At the top of the two towers, dot product of these two vectors from both sides is computed, representing similarity between seed users and target user. As to recommendation, the similarity is essentially the possibility of specific item being clicked by the target user. 

\textbf{\textit{Transforming matrix}}. This weight matrix sized $m\times{h}$ is designed to make projection from universal user embedding vector space to a look-alike aware space. Although user embeddings are learned from variety kinds of behavior, taking pre-trained features as input can still possibly cause overfitting. To prevent overfitting, we make both towers share the training of transforming matrix. ReLU unit is used before output to catch nonlinear features. After transforming, the group of seed users are represented as $n$ vectors of size $h$. 

\begin{figure}[h]
    \centering
    \includegraphics[scale=0.35]{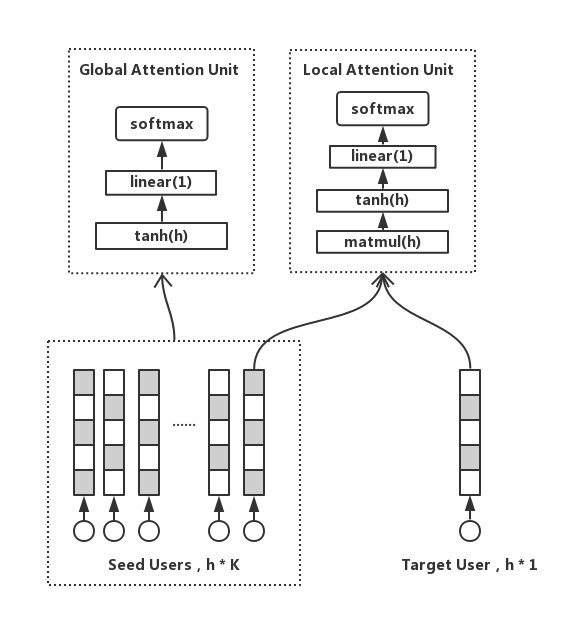}
    \caption{Local and global attention units.}
    \label{lookalike attention}
\end{figure}

\textbf{\textit{Local attention}}. To compute similarity between seeds and the target user, we have to pooling seeds vectors into one. Average pooling is a commonly used method. However, averaging is to find the centroid of seeds, leading to a result that common information is saved whereas outliers and personalize information are m. In general, among millions of seed users, there might be only a small part of users whose interest is related to the target user. Thus we pose a local attention unit here to activate local interest with regard to a certain target users and adaptively learn a personalized representation of seeds to the target user:
\begin{equation}E_{local_{s}}=E_{s}softmax(\tanh(E_{s}^{T}W_{l}E_{u})),\end{equation}
where $W_{l}\in{\mathbb{R}^{h\times{h}}}$ denotes the attention matrix, $E_s$ denotes seed users, $E_u$ denotes target user, $E_{local_{s}}$ is an embedding of local information of seeds. Notice that if an item has millions of seed users, it will cost $n*h*h$ times computing for local attention where $n$ is million level, causing problem in online predicting. To reduce the time complexity of computation, we cluster the seed users through K-means algorithm into $k$ clusters, and for each cluster we calculate the average mean of seeds vectors. By doing this, we get $k$ vectors of size $h$, where $k$ represents the number of clusters. Thus calculation times is reduced from $n*h*h$ to $k*h*h$, where $k$ is usually less than 100. Notice that K-means clustering is taken before epoch starts and cluster centroids are saved in memory, so that only $n$ times of inference are calculated during each epoch of training. 

\textbf{\textit{Global attention}}. As for global information of seed users, we add a self attention unit to model the weight each seeds cluster contribute to the global representation:
\begin{equation}E_{global_{s}}=E_{s}softmax(E_{s}^{T}\tanh(W_{g}E_{s})),\end{equation}
where $W_{g}\in{\mathbb{R}^{s\times{n}}}$ denotes the attention matrix and $s$ denotes the attention size, $E_{global_{s}}$ outputs a weighted summed vector representing the global information of seed users $E_{s}$. The global information extracted by self attention is related to the interest distribution of $E_{s}$ themselves. As Fig.4, with both local and global information $E_{global}$ and $E_{local}$, we can calculate the similarity between seeds and target user according to Eq.(1). 

\textbf{\textit{Iterative training}}. After transformation and back propagation, values of user embeddings are changed. In order to keep feeds clustering synchronous with user embeddings, we rerun feeds clustering after each epoch ends. So we propose an iterative training process that train look-alike learning model and clustering alternately round by round.

\textbf{\textit{Loss}}. We use sigmoid cross entropy function as loss function:
\begin{equation}L=-\frac{1}{N}\sum_{x,y\in{D}}(y\log{p(x)} + (1-y)\log(1-p(x))),\end{equation}
where D represents the training set, x denotes user embedding as input, y is the label in $\{0,1\}$, $p(x)$ is the prediction score between seed users and target user through sigmoid function.

\section{EXPERIMENTS}

\subsection{Experimental Setup}
\textbf{Dataset}. We use real traffic logs from Wechat "Top Stories" recommender system, which are randomly sampled and contains 2,984,724 users, 43,500 articles and 8,891,670 samples. Features include user gender, age, interested categories and tags and historically read articleID. The samples are formatted as $\{uid,item\_id,is\_click\}$. Notice that in order to make model be fair, we eliminate users who read more than 50 articles or have no behaviors a day. While training, user features and seed user ids of item is joined by $uid$ and $item\_id$. The model takes all features as input and outputs predicting score for each user-item pair.

\textbf{Model setup}. We use Adam optimizer with learning rate of 0.001. The mini-batch size is 256. The embedding size of both user representation and look-alike embeddings is set to 128. The number of clusters $k$ is set to 20.

\subsection{Competitors}
\begin{itemize}
\item \textbf{LR}. Logistic Regression is a commonly used regression-based look-alike model. We implement it to verify effectiveness of deep model.
\item \textbf{Yahoo Look-alike Model}. A model combining similarity-based and regression-based method. To effectively compare, we choose information value as feature selection, and take $p_j$ > 0.5 as positive filter. 
\item \textbf{Youtube DNN}. Youtube DNN is a deep model using concatenation merging multi-field user features, whereas RALM uses attention merge layer. We implement it to compare expressive ability of user interests between these two models.
\item \textbf{RALM with average}. To represent seed users, averaging is a simple way of pooling multiple user emebeddings. To prove effectiveness of attention mechanism on global and local information, we set a model using average pooling to merge all seed users to one, instead of attention units.  
\end{itemize}

\subsection{Metrics}
Generally, loss and AUC are commonly used metrics in ranking model evaluation. However, the competing models define different loss function and make it hard to compare loss. Moreover, AUC is related to positive/negative proportion in samples, and sometimes cannot truly reflect online performance. We introduce a metric called "precision at K", marked as prec@K, denoting how many of recommended top K results will hit user's actually read items. The equation is as follows:
\begin{equation}prec@K=\frac{1}{N}\sum_{i}^{N}\frac{size(R_{iK} \cap S_{i})}{min(K,size(S_{i}))},\end{equation}
where $K$ denotes number of recommended results, $R_{iK}$ denotes top $K$ items in user $i$'s recommended results ordered decreasingly by predicting score, $S$ denotes item set that user has read, $N$ represents number of users in test set. Notice that some users may read less than K items, leading to a irregular low prec@K, so we get a minimum between $K$ and size of $S_{i}$. In this experiment, we compare models by AUC and prec@K with $K\in\{10,50\}$.

\begin{table}[htbp]
	\centering
	\caption{Model comparison on Wechat Dataset.}
	\begin{tabular}{llll}
		\toprule
		Model&AUC&prec@10&prec@50 \\ 
		\midrule
        LR&0.5252&0.0811&0.0729 \\
		Yahoo Look-alike Model&0.5512&0.1023&0.0941 \\
        Youtube DNN&0.5903&0.1217&0.1087 \\
        \textbf{RALM with average}&\textbf{0.5842}&\textbf{0.1108}&\textbf{0.0980} \\
        \textbf{RALM}&\textbf{0.6012}&\textbf{0.1295}&\textbf{0.1099}\\
		\bottomrule
	\end{tabular}
\end{table}

\subsection{Results from offline evaluation}
Table 1 shows the experimental results on Wechat Dataset. Notice that our dataset is large, and in order to make comparison quickly, we take a few features into consideration. As a result, AUC is not as high as that on public datasets. 

We can see LR performs not well, verifying advantages of deep models. Besides it is due to its shortage in capturing information of clusters. Yahoo Look-alike Model performs better than LR. Meanwhile, it is weak than deep models. The reason is that it is posed to use in massive case, and take a non-iterative feature selection as filter. Youtube DNN achieves 0.0391 absolute AUC gain over Yahoo Look-alike Model, proving effectiveness of deep model. RALM with average pooling achieves 0.033 absolute AUC gain over Yahoo Look-alike Model, but is weak than Youtube DNN. It implies that simply averaging user embeddings is not as effective as an end-to-end deep model. RALM with attention units performs the best among all the models. It achieves 0.0109 absolute AUC gain and 0.0078/0.0022 prec@10/prec@50 gain over Youtube DNN. This improvement verifies that attention mechanism does help in extracting seed users' information and find local relation between seeds and target users. 
\begin{figure}[h]
    \centering
    \includegraphics[scale=0.213]{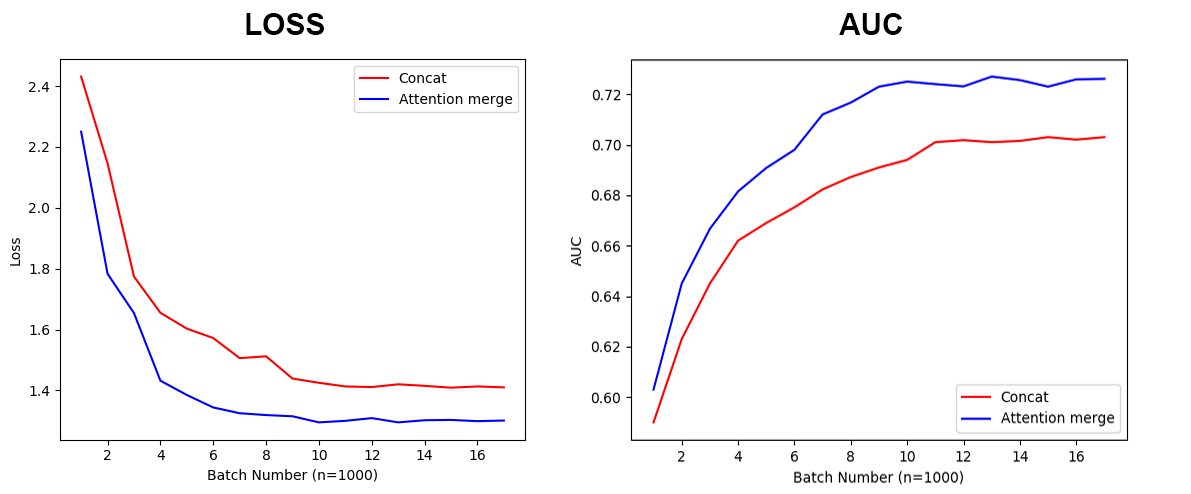}
    \caption{Performances of user representation learning model with concatenation and attention merge.}
    \label{representation loss}
\end{figure}
\subsection{Experiments on attention merge layer}
In user representation learning, we have many individual fields of feature to add into deep model. In order to learn relations and interactions among different fields, we make a comparison between concatenation and attention merge layer in user representation learning. 

Shown as Fig.5, attention merge layer performs much better than concatenation in AUC and loss in testing set, and spend less batches to converge. We owe it to the design of attention unit. When fields are concatenated and fed forward, same parts of neurons are activated to all users. When it comes to attention merge layer, different neurons will be activated faced to different users, meaning that personalize interactions among features are learned, which do a favor for the model performance.

\subsection{Experiments on clustering}
In look-alike learning, cluster number $k$ is a key parameter in K-means clustering. We conduct an experiment to observe how $k$ value affects the performance. In this experiment, $k$ is set to 5,10,20,50,100, and each metric is averaged from testing set after 5 epochs training.
\begin{figure}[h]
    \centering
    \includegraphics[scale=0.213]{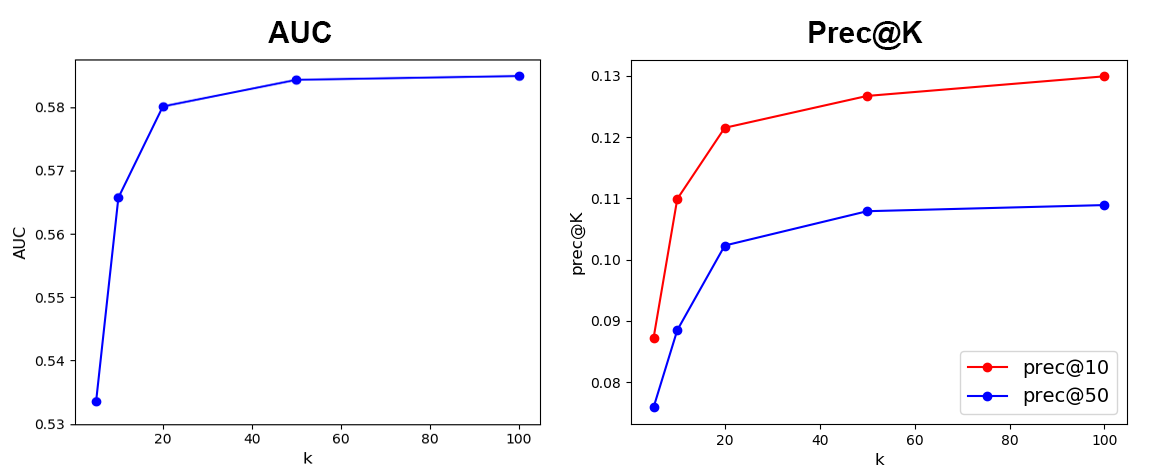}
    \caption{Performances of RALM with different k values. After $k=20$, AUC and prec@K are stable.}
    \label{k experiment}
\end{figure}

Fig.6 shows that as $k$ increases, the performance of model gets better. It indicates that the more clusters generated, the less information of seeds lost. On the other hand, bigger $k$ means more computation in online prediction. In Fig.6 we can see an elbow point in curves where $k=20$, after which the metrics hardly get higher. As a compromise, we set $k=20$ in our online model.  

\begin{table}[htbp]
	\centering
	\caption{Relative changes in metrics compared to control group in online A/B testing.}
	\begin{tabular}{ll}
    	\toprule
		Metric&Change\\ 
		\midrule
		Exposure&+9.112\%\\
        CTR&+1.09\%\\
        Category Diversity&+8.435\%\\
        Tag Diversity&+15.938\%\\
        GINI Coefficient&+5.36\%\\
		\bottomrule
	\end{tabular}
\end{table}
\subsection{Results from online A/B testing}
To verify the real benefits RALM brings to our recommender system, we conduct an online A/B testing in Wechat "Top Stories" recommender system from 2018-11 to 2018-12. We split online traffic by userID and arrange the same proportion to control group and experimental group. To evaluate the performance online, following metrics are involved:

\textbf{Exposure}: An exposure means a user read an item, and it is only counted once for a user. As audience extended by look-alike model, more user should be reached for given items, and exposure is expected to increase.

\textbf{CTR (Click-through Rate)}: As audience increased, many new users sharing the same interests with seeds are reached. Therefore, CTR is expected not to decrease.

\textbf{Category\&Tag Diversity}. One of our purpose is enriching user's interest in our system, so we define a metric named diversity. It is represented by number of content categories or tags a user has read in a day. With a more comprehensive user representation, more kinds of contents will be reached and category\&tag diversity are expected to increase. 

\textbf{Gini Coefficient}. RALM aims at relieve Mathew Effect, thus we use Gini coefficient to measure the distribution of clicks on all items in recommender system. A higher Gini coefficient indicates that more long-tail contents are consumed and the system is much better at distributing contents.

From Table 2 we can see owing to audience extension, exposure gains a great promotion. Meanwhile CTR increases a little, showing that the extended users are exactly interested in those contents as well. Furthermore, by applying attention merge layer in user representation, more hidden factors affecting user's interest are mined. As a result, audiences are able to be extended via more aspects of interest, and category\&tag diversity are increased significantly. In general, RALM makes a high quality and diverse extension for the seeds, by which all recommended contents can reach proper target users. Correspondingly, Gini coefficient achieves a 5.36\% gain.

\section{CONCLUSIONS}
In this paper, we present RALM as a way to realize real time audience extension. We propose a two-phase modeling in RALM including user representation learning and look-alike learning. In user representation learning, we introduce attention merge to learn effective relations among different fields of user features. In look-alike learning, we design local and global attention unit to learn an adaptive representation of seed users with respect to target user. Based on this model, we build a system containing offline training and online serving. With techniques such as asynchronous preprocessing and seeds clustering, online predicting is able to be real-time. Now RALM based audience extension system has been deployed in Wechat "Top Stories" recommender system.

%
\bibliographystyle{ACM-Reference-Format}

\bibliography{ralm}


\begin{thebibliography}{19}


\ifx \showCODEN    \undefined \def \showCODEN     #1{\unskip}     \fi
\ifx \showDOI      \undefined \def \showDOI       #1{#1}\fi
\ifx \showISBNx    \undefined \def \showISBNx     #1{\unskip}     \fi
\ifx \showISBNxiii \undefined \def \showISBNxiii  #1{\unskip}     \fi
\ifx \showISSN     \undefined \def \showISSN      #1{\unskip}     \fi
\ifx \showLCCN     \undefined \def \showLCCN      #1{\unskip}     \fi
\ifx \shownote     \undefined \def \shownote      #1{#1}          \fi
\ifx \showarticletitle \undefined \def \showarticletitle #1{#1}   \fi
\ifx \showURL      \undefined \def \showURL       {\relax}        \fi
\providecommand\bibfield[2]{#2}
\providecommand\bibinfo[2]{#2}
\providecommand\natexlab[1]{#1}
\providecommand\showeprint[2][]{arXiv:#2}

\bibitem[\protect\citeauthoryear{Bahdanau, Cho, and Bengio}{Bahdanau
  et~al\mbox{.}}{2014}]%
        {bahdanau2014neural}
\bibfield{author}{\bibinfo{person}{Dzmitry Bahdanau},
  \bibinfo{person}{Kyunghyun Cho}, {and} \bibinfo{person}{Yoshua Bengio}.}
  \bibinfo{year}{2014}\natexlab{}.
\newblock \showarticletitle{Neural machine translation by jointly learning to
  align and translate}.
\newblock \bibinfo{journal}{\emph{arXiv preprint arXiv:1409.0473}}
  (\bibinfo{year}{2014}).
\newblock


\bibitem[\protect\citeauthoryear{Chen, Zhang, He, Nie, Liu, and Chua}{Chen
  et~al\mbox{.}}{2017}]%
        {chen2017attentive}
\bibfield{author}{\bibinfo{person}{Jingyuan Chen}, \bibinfo{person}{Hanwang
  Zhang}, \bibinfo{person}{Xiangnan He}, \bibinfo{person}{Liqiang Nie},
  \bibinfo{person}{Wei Liu}, {and} \bibinfo{person}{Tat-Seng Chua}.}
  \bibinfo{year}{2017}\natexlab{}.
\newblock \showarticletitle{Attentive collaborative filtering: Multimedia
  recommendation with item-and component-level attention}. In
  \bibinfo{booktitle}{\emph{Proceedings of the 40th International ACM SIGIR
  conference on Research and Development in Information Retrieval}}. ACM,
  \bibinfo{pages}{335--344}.
\newblock


\bibitem[\protect\citeauthoryear{Cheng, Koc, Harmsen, Shaked, Chandra, Aradhye,
  Anderson, Corrado, Chai, Ispir, et~al\mbox{.}}{Cheng et~al\mbox{.}}{2016}]%
        {cheng2016wide}
\bibfield{author}{\bibinfo{person}{Heng-Tze Cheng}, \bibinfo{person}{Levent
  Koc}, \bibinfo{person}{Jeremiah Harmsen}, \bibinfo{person}{Tal Shaked},
  \bibinfo{person}{Tushar Chandra}, \bibinfo{person}{Hrishi Aradhye},
  \bibinfo{person}{Glen Anderson}, \bibinfo{person}{Greg Corrado},
  \bibinfo{person}{Wei Chai}, \bibinfo{person}{Mustafa Ispir}, {et~al\mbox{.}}}
  \bibinfo{year}{2016}\natexlab{}.
\newblock \showarticletitle{Wide \& deep learning for recommender systems}. In
  \bibinfo{booktitle}{\emph{Proceedings of the 1st Workshop on Deep Learning
  for Recommender Systems}}. ACM, \bibinfo{pages}{7--10}.
\newblock


\bibitem[\protect\citeauthoryear{Covington, Adams, and Sargin}{Covington
  et~al\mbox{.}}{2016}]%
        {covington2016deep}
\bibfield{author}{\bibinfo{person}{Paul Covington}, \bibinfo{person}{Jay
  Adams}, {and} \bibinfo{person}{Emre Sargin}.}
  \bibinfo{year}{2016}\natexlab{}.
\newblock \showarticletitle{Deep neural networks for youtube recommendations}.
  In \bibinfo{booktitle}{\emph{Proceedings of the 10th ACM Conference on
  Recommender Systems}}. ACM, \bibinfo{pages}{191--198}.
\newblock


\bibitem[\protect\citeauthoryear{Hartigan and Wong}{Hartigan and Wong}{1979}]%
        {hartigan1979algorithm}
\bibfield{author}{\bibinfo{person}{John~A Hartigan} {and}
  \bibinfo{person}{Manchek~A Wong}.} \bibinfo{year}{1979}\natexlab{}.
\newblock \showarticletitle{Algorithm AS 136: A k-means clustering algorithm}.
\newblock \bibinfo{journal}{\emph{Journal of the Royal Statistical Society.
  Series C (Applied Statistics)}} \bibinfo{volume}{28}, \bibinfo{number}{1}
  (\bibinfo{year}{1979}), \bibinfo{pages}{100--108}.
\newblock


\bibitem[\protect\citeauthoryear{Kim and Kim}{Kim and Kim}{2006}]%
        {kim2006performance}
\bibfield{author}{\bibinfo{person}{Myung~Won Kim} {and} \bibinfo{person}{Eun~Ju
  Kim}.} \bibinfo{year}{2006}\natexlab{}.
\newblock \showarticletitle{Performance improvement in collaborative
  recommendation using multi-layer perceptron}. In
  \bibinfo{booktitle}{\emph{International Conference on Neural Information
  Processing}}. Springer, \bibinfo{pages}{350--359}.
\newblock


\bibitem[\protect\citeauthoryear{Lin, Feng, Santos, Yu, Xiang, Zhou, and
  Bengio}{Lin et~al\mbox{.}}{2017}]%
        {lin2017structured}
\bibfield{author}{\bibinfo{person}{Zhouhan Lin}, \bibinfo{person}{Minwei Feng},
  \bibinfo{person}{Cicero Nogueira~dos Santos}, \bibinfo{person}{Mo Yu},
  \bibinfo{person}{Bing Xiang}, \bibinfo{person}{Bowen Zhou}, {and}
  \bibinfo{person}{Yoshua Bengio}.} \bibinfo{year}{2017}\natexlab{}.
\newblock \showarticletitle{A structured self-attentive sentence embedding}.
\newblock \bibinfo{journal}{\emph{arXiv preprint arXiv:1703.03130}}
  (\bibinfo{year}{2017}).
\newblock


\bibitem[\protect\citeauthoryear{Liu, Pardoe, Liu, Thakur, Cao, and Li}{Liu
  et~al\mbox{.}}{2016}]%
        {liu2016audience}
\bibfield{author}{\bibinfo{person}{Haishan Liu}, \bibinfo{person}{David
  Pardoe}, \bibinfo{person}{Kun Liu}, \bibinfo{person}{Manoj Thakur},
  \bibinfo{person}{Frank Cao}, {and} \bibinfo{person}{Chongzhe Li}.}
  \bibinfo{year}{2016}\natexlab{}.
\newblock \showarticletitle{Audience expansion for online social network
  advertising}. In \bibinfo{booktitle}{\emph{Proceedings of the 22nd ACM SIGKDD
  International Conference on Knowledge Discovery and Data Mining}}. ACM,
  \bibinfo{pages}{165--174}.
\newblock


\bibitem[\protect\citeauthoryear{Ma, Wen, Xia, and Chen}{Ma
  et~al\mbox{.}}{2016}]%
        {ma2016sub}
\bibfield{author}{\bibinfo{person}{Qiang Ma}, \bibinfo{person}{Musen Wen},
  \bibinfo{person}{Zhen Xia}, {and} \bibinfo{person}{Datong Chen}.}
  \bibinfo{year}{2016}\natexlab{}.
\newblock \showarticletitle{A Sub-linear, Massive-scale Look-alike Audience
  Extension System A Massive-scale Look-alike Audience Extension}. In
  \bibinfo{booktitle}{\emph{Workshop on Big Data, Streams and Heterogeneous
  Source Mining: Algorithms, Systems, Programming Models and Applications}}.
  \bibinfo{pages}{51--67}.
\newblock


\bibitem[\protect\citeauthoryear{Mangalampalli, Ratnaparkhi, Hatch,
  Bagherjeiran, Parekh, and Pudi}{Mangalampalli et~al\mbox{.}}{2011}]%
        {mangalampalli2011feature}
\bibfield{author}{\bibinfo{person}{Ashish Mangalampalli},
  \bibinfo{person}{Adwait Ratnaparkhi}, \bibinfo{person}{Andrew~O Hatch},
  \bibinfo{person}{Abraham Bagherjeiran}, \bibinfo{person}{Rajesh Parekh},
  {and} \bibinfo{person}{Vikram Pudi}.} \bibinfo{year}{2011}\natexlab{}.
\newblock \showarticletitle{A feature-pair-based associative classification
  approach to look-alike modeling for conversion-oriented user-targeting in
  tail campaigns}. In \bibinfo{booktitle}{\emph{Proceedings of the 20th
  international conference companion on World wide web}}. ACM,
  \bibinfo{pages}{85--86}.
\newblock


\bibitem[\protect\citeauthoryear{Mikolov, Sutskever, Chen, Corrado, and
  Dean}{Mikolov et~al\mbox{.}}{2013}]%
        {mikolov2013distributed}
\bibfield{author}{\bibinfo{person}{Tomas Mikolov}, \bibinfo{person}{Ilya
  Sutskever}, \bibinfo{person}{Kai Chen}, \bibinfo{person}{Greg~S Corrado},
  {and} \bibinfo{person}{Jeff Dean}.} \bibinfo{year}{2013}\natexlab{}.
\newblock \showarticletitle{Distributed representations of words and phrases
  and their compositionality}. In \bibinfo{booktitle}{\emph{Advances in neural
  information processing systems}}. \bibinfo{pages}{3111--3119}.
\newblock


\bibitem[\protect\citeauthoryear{Pazzani and Billsus}{Pazzani and
  Billsus}{2007}]%
        {pazzani2007content}
\bibfield{author}{\bibinfo{person}{Michael~J Pazzani} {and}
  \bibinfo{person}{Daniel Billsus}.} \bibinfo{year}{2007}\natexlab{}.
\newblock \showarticletitle{Content-based recommendation systems}.
\newblock In \bibinfo{booktitle}{\emph{The adaptive web}}.
  \bibinfo{publisher}{Springer}, \bibinfo{pages}{325--341}.
\newblock


\bibitem[\protect\citeauthoryear{Qu, Wang, Sun, and Holtan}{Qu
  et~al\mbox{.}}{2014}]%
        {qu2014systems}
\bibfield{author}{\bibinfo{person}{Yan Qu}, \bibinfo{person}{Jing Wang},
  \bibinfo{person}{Yang Sun}, {and} \bibinfo{person}{Hans~Marius Holtan}.}
  \bibinfo{year}{2014}\natexlab{}.
\newblock \bibinfo{title}{Systems and methods for generating expanded user
  segments}.
\newblock
\newblock
\newblock
\shownote{US Patent 8,655,695.}


\bibitem[\protect\citeauthoryear{Rendle}{Rendle}{2010}]%
        {rendle2010factorization}
\bibfield{author}{\bibinfo{person}{Steffen Rendle}.}
  \bibinfo{year}{2010}\natexlab{}.
\newblock \showarticletitle{Factorization machines}. In
  \bibinfo{booktitle}{\emph{Data Mining (ICDM), 2010 IEEE 10th International
  Conference on}}. IEEE, \bibinfo{pages}{995--1000}.
\newblock


\bibitem[\protect\citeauthoryear{Sarwar, Karypis, Konstan, and Riedl}{Sarwar
  et~al\mbox{.}}{2001}]%
        {sarwar2001item}
\bibfield{author}{\bibinfo{person}{Badrul Sarwar}, \bibinfo{person}{George
  Karypis}, \bibinfo{person}{Joseph Konstan}, {and} \bibinfo{person}{John
  Riedl}.} \bibinfo{year}{2001}\natexlab{}.
\newblock \showarticletitle{Item-based collaborative filtering recommendation
  algorithms}. In \bibinfo{booktitle}{\emph{Proceedings of the 10th
  international conference on World Wide Web}}. ACM, \bibinfo{pages}{285--295}.
\newblock


\bibitem[\protect\citeauthoryear{Shen, Geyik, and Dasdan}{Shen
  et~al\mbox{.}}{2015}]%
        {shen2015effective}
\bibfield{author}{\bibinfo{person}{Jianqiang Shen}, \bibinfo{person}{Sahin~Cem
  Geyik}, {and} \bibinfo{person}{Ali Dasdan}.} \bibinfo{year}{2015}\natexlab{}.
\newblock \showarticletitle{Effective audience extension in online
  advertising}. In \bibinfo{booktitle}{\emph{Proceedings of the 21th ACM SIGKDD
  International Conference on Knowledge Discovery and Data Mining}}. ACM,
  \bibinfo{pages}{2099--2108}.
\newblock


\bibitem[\protect\citeauthoryear{Tufekci}{Tufekci}{2008}]%
        {tufekci2008can}
\bibfield{author}{\bibinfo{person}{Zeynep Tufekci}.}
  \bibinfo{year}{2008}\natexlab{}.
\newblock \showarticletitle{Can you see me now? Audience and disclosure
  regulation in online social network sites}.
\newblock \bibinfo{journal}{\emph{Bulletin of Science, Technology \& Society}}
  \bibinfo{volume}{28}, \bibinfo{number}{1} (\bibinfo{year}{2008}),
  \bibinfo{pages}{20--36}.
\newblock


\bibitem[\protect\citeauthoryear{Vaswani, Shazeer, Parmar, Uszkoreit, Jones,
  Gomez, Kaiser, and Polosukhin}{Vaswani et~al\mbox{.}}{2017}]%
        {vaswani2017attention}
\bibfield{author}{\bibinfo{person}{Ashish Vaswani}, \bibinfo{person}{Noam
  Shazeer}, \bibinfo{person}{Niki Parmar}, \bibinfo{person}{Jakob Uszkoreit},
  \bibinfo{person}{Llion Jones}, \bibinfo{person}{Aidan~N Gomez},
  \bibinfo{person}{{\L}ukasz Kaiser}, {and} \bibinfo{person}{Illia
  Polosukhin}.} \bibinfo{year}{2017}\natexlab{}.
\newblock \showarticletitle{Attention is all you need}. In
  \bibinfo{booktitle}{\emph{Advances in Neural Information Processing
  Systems}}. \bibinfo{pages}{5998--6008}.
\newblock


\bibitem[\protect\citeauthoryear{Zhou, Zhu, Song, Fan, Zhu, Ma, Yan, Jin, Li,
  and Gai}{Zhou et~al\mbox{.}}{2018}]%
        {zhou2018deep}
\bibfield{author}{\bibinfo{person}{Guorui Zhou}, \bibinfo{person}{Xiaoqiang
  Zhu}, \bibinfo{person}{Chenru Song}, \bibinfo{person}{Ying Fan},
  \bibinfo{person}{Han Zhu}, \bibinfo{person}{Xiao Ma},
  \bibinfo{person}{Yanghui Yan}, \bibinfo{person}{Junqi Jin},
  \bibinfo{person}{Han Li}, {and} \bibinfo{person}{Kun Gai}.}
  \bibinfo{year}{2018}\natexlab{}.
\newblock \showarticletitle{Deep interest network for click-through rate
  prediction}. In \bibinfo{booktitle}{\emph{Proceedings of the 24th ACM SIGKDD
  International Conference on Knowledge Discovery \& Data Mining}}. ACM,
  \bibinfo{pages}{1059--1068}.
\newblock


\end{thebibliography}

%
\appendix

\end{document}